\documentstyle[preprint,eqsecnum,aps,epsfig,rotating]{revtex}

\begin{document}

\draft
\newcommand\ie {{\it i.e.}}
\newcommand\eg {{\it e.g.}}
\newcommand\etc{{\it etc.}}
\newcommand\etal {{\it et al. }}
\newcommand{\be}{\begin{eqnarray}}
\newcommand{\ee}{\end{eqnarray}}
\newcommand{\bem}{\begin{mathletters}}
\newcommand{\eem}{\end{mathletters}}
\newcommand{\intk}{{i\over \beta} \sum_n e^{i\omega_n \eta} 
\int^\Lambda {d^3k \over (2\pi)^3}}
%\newcommand\noi {\noindent}

%\preprint{HD-TVP-97/XX}
%\preprint{}

\title{$\pi\pi$ scattering in the $\rho$-meson channel
at finite temperature}

\author{Y.~B.~ He, J.~H\"ufner, S.~P.~Klevansky, and P.~Rehberg }
\address{Institut f\"ur Theoretische Physik, \\
Philosophenweg 19, D-69120 Heidelberg, Germany}

%\date{\today}

\maketitle
\vspace{1cm}

\begin{abstract}
We study $\pi\pi$ scattering in the $I=1$, $J^P=1^-$ channel
at finite temperature 
in the framework of the extended 
Nambu-Jona-Lasinio model that explicitly includes vector and axial-vector
degrees of freedom in addition to the usual scalar and  pseudoscalar 
sector. The $S$-matrix in the coupled channels $q\bar q$ and $\pi
\pi$ is constructed via $\rho$-exchange in the $s$-channel.  
The self-energy of the $\rho$-meson contains both quark and pion loop 
contributions. The analytic structure of the $S$-matrix for $T\geq 0$ is
investigated and the motion of the $\rho$-pole as a function of coupling
constant and temperature is followed in the 
complex $\sqrt{s}$-plane. 
For numerical calculations, 
parameters are chosen
in order that $m_\pi$, $f_\pi$ and the
 experimental $\pi\pi$  phase shifts $\delta_1^1$ at zero
temperature are reproduced, and then the 
behavior of the $\rho$-pole as well as 
 the $\pi\pi$ cross section is investigated
 as a function of the temperature.  We find that
 the position of the $\rho$ mass stays practically constant for 
$0\leq T\leq 130$~MeV, and  then
moves down in energy by about 200~MeV for 130~MeV$\leq T\leq 230$~MeV.
\end{abstract}
\pacs{25.75.-q, 12.39.Ba, 21.65.+f, 24.85.+p
\\ {\bf Keywords}: $\pi\pi$ scattering, $\rho$-meson, finite temperature,
analytic structure.}
\narrowtext

\newpage
\section{Introduction}
During the last decade many experimental and 
theoretical efforts have been dedicated to the study
of relativistic heavy-ion 
collisions, in which one hopes to observe 
the two aspects of the 
 QCD phase transition: (i) the confinement / deconfinement transition
in which 
 hadronic matter decomposes into the quark-gluon plasma, and the converse
process, and  
(ii) that of chiral symmetry restoration / breaking, both of which should
occur at high values of the  temperature
and / or density. Depending on the energy density
reached, 
the hot fireball formed after the collisions
may (or may not) enter 
 into the quark-gluon plasma phase and 
undergo a chiral transition simultaneously before it 
cools down and 
hadronizes, leaving it a challenge for 
people to identify its trajectory from the observed hadronic 
products. In particular, during the evolution of the fireball,
a large number of pions is produced in the mid-rapidity region, 
and one can expect that interactions between them
play an essential role in the dynamical evolution.
It is thus of  interest to study the $\pi\pi$ scattering in a
thermal environment in order to understand the evolution dynamics of
the system.
The $\pi\pi$ scattering cross section, for example,  is an important
input to transport equations that are used to simulate the dynamical
evolution of the fireball. 
Based on hadronic models,
Refs.~\cite{hwBar92,vMul92} have investigated the $\pi\pi$ cross section
in nuclear matter, and have claimed to observe
 considerable medium effects.   In itself,
a study of the temperature dependence of the $\pi\pi$ scattering cross
section is necessary, and we will examine this here in the framework of
a chiral model.
On the other hand, $\pi\pi$ annihilation is 
one important source of dilepton production. 
In recent studies by 
the CERES and HELIOS-3 collaborations \cite{ceres95},
an enhancement of the dilepton spectrum has been observed in the energy 
range of 200 - 800 MeV in nucleus-nucleus collisions.
One of the hypotheses brought forward to explain these data is that
the $\rho$-meson mass shifts to lower values as a function of increasing
temperature
and density, and its width broadens while doing so \cite{ko}.    
Since a major source of dileptons in this energy range comes from the
$\rho$-meson, it is necessary to understand
to what extent a mass shift can occur, and if one occurs, whether it is
an increase or decrease.   Thus, via our study of the $\pi\pi$ system, we
also wish to examine the medium dependence of the $\rho$-meson. 
These two problems, that of $\pi\pi$ scattering at energies 
500~MeV$<\sqrt{s} <$900 MeV and that of the medium 
dependence of the $\rho$-meson, are closely intertwined, and we address
both of these problems 
in this paper.   

 The medium effects of the $\rho$-meson have been 
extensively discussed recently and for a review we refer the 
reader to
Ref.~\cite{Ko97}.  
The $\rho$-meson properties at finite temperature are not yet well
understood, and in particular the issue as to 
whether the $\rho$ mass increases or decreases with 
increasing temperature 
has not yet been settled \cite{gal91,bro91,ele95,pis95}.
In Ref.~\cite{gal91}, for example, a vector dominance model that 
does not support a chiral or deconfinement phase transition is used 
to describe the temperature dependence,
and these authors find a weak rising temperature dependence of the 
$\rho$ mass and width.
On the other hand, 
recent lattice calculations have found no sign of a temperature
dependence in the meson properties for temperatures up to 0.92$T_c$ 
\cite{boy95}.
In this work, we study the temperature dependence within
 a chiral model that displays a phase transition in the chiral
limit and a smooth crossover for small values of the current quark mass.
We investigate the analytic
structure of the $\rho$ propagator and consider the $\rho$-meson as
a pole in the complex plane. In this way we contribute additional insight.

Our starting point is an  effective chiral model:   we use
a version of the Nambu-Jona-Lasinio (NJL) model \cite{NJL,NJLrev}
that has been extended to explicitly include vector and axial-vector degrees
of freedom \cite{sKli90,bijnens,lemmer}
in order to study the $\pi\pi$ cross section 
at finite temperature and vanishing chemical potential under inclusion of
the $\rho$-meson.
The NJL-model, constructed essentially on the observation of 
the  chiral symmetry of quantum chromodynamics (QCD)
gives  a transparent description of 
the mechanism of spontaneous
symmetry
breaking.   It  has been extensively used to study the static meson properties,
and has enjoyed remarkable success, especially in the 
pseudoscalar sector
\cite{NJL,NJLrev}.
However,
to describe the  $\rho$-meson in the context of the
NJL-model is a delicate task \cite{ahBli90,mTak91},
since, on heuristic grounds,  $m_\rho\simeq 2m_q$, where
$m_q\sim 300$-400~MeV is the mass of the constituent quark,
and the NJL-model does not confine quarks.
In the language of scattering theory, the $\rho$ mass is close to the
$q\bar q$ threshold.
While theshold effects often modify the shape of a resonance on the 
real energy axis close by, the position of a pole in the complex plane
is much less distorted. For this reason we choose to define the properties of
the $\rho$-meson as those of the pole in the complex plane.
The self-energy of the $\rho$-meson in the
NJL-model arises from the quark loop 
contribution at the lowest order in an expansion in 
the inverse number of colors, $1/N_c$,
 which leads to the unphysical decay $\rho \rightarrow 
q\bar q$ that is due  to the lack of confinement.
\footnote{In spite of this deficiency, this model still appears to
give a good description of the position of the $\rho$-mass 
\cite{ber97,lemmer}. }
On the other hand, it is known from 
experiment that almost all $\rho$-mesons decay via the channel
$\rho\rightarrow \pi\pi$,
and thus a satisfactory description of the $\rho$-meson
cannot be obtained unless the physical $\rho$-meson
decay channel has been taken into
account.
From the point of view of the NJL model, the necessary pion loop corrections
can be classified as part of the next order in the $1/N_c$
expansion \cite{quack,dmitra}.   In fact, these terms should represent
the leading contributions at this order \cite{hippe}.   It is also
necessary 
 to include the pion loop contribution
to the $\rho$-meson self-energy in order to fulfill the unitarity of the
$\pi\pi$ scattering amplitude. With the inclusion of both pion loop and quark
loop contributions to the $\rho$-meson self-energy, the $\pi\pi$ scattering
via $\rho$-meson exchange becomes a coupled channel (\ie \   $\pi\pi$ and 
$q\bar q$ channels) process.
Care has to be taken to obey current conservation.
We study the analytic structure of the 
$\rho$-meson propagator or
 scattering amplitude in the complex energy 
plane, in order to identify the $\rho$ 
mass as a  pole. This part of the work is completely new.
The parameters of the model
can be chosen such that at $T=0$, the values of $m_\pi$, $f_\pi$ and
the experimental data of the $\pi\pi$ phase shifts are well reproduced.
Temperature effects are incorporated via the 
imaginary time formalism and the temperature 
dependence of the $\pi\pi$ cross section is obtained.
Note that $\pi\pi$ scattering has been studied by several authors in the
context of the NJL model, however only to obtain the scattering lengths
at zero temperature  without the inclusion of the vector meson
sector\cite{rui94},
and also with it \cite{party}.   
Scattering lengths at finite
temperature were calculated in \cite{qua95}.

One may well ask the question as to the validity of the NJL model 
in describing the $\rho$-meson in the
manner described. No conceptual difficulty arises 
with the cutoff. It restricts the
three momentum $\left| \vec{p} \right | <\Lambda$ and
these is no condition on the mass. 
The problem lies rather in the fact that the unphysical $\rho\rightarrow
q\bar q$ channel is open, and the poles of the propagator, if
taken only in the 
lowest order in the $1/N_c$ expansion, will only reflect this physics.
In spite of this, several authors \cite{ber97,lemmer} have found that the 
static properties in the vector meson sector can still be well understood,
and that this model can still be used in a range which may 
{\it a priori} seem to
lie outside of its realm of validity. Although we have gone beyond this 
restrictive approximation allowing in addition the physical decay channel
$\rho\rightarrow \pi\pi$ to open, we are still able to observe and 
isolate the effects of the opening of the 
unphysical $q\bar q $ threshold. This can be
clearly seen in the calculated $\pi\pi$ cross section, and is discussed 
at that point.

This paper is organized as follows. In Sec.~\ref{sect:NJL},
we describe our formalism in the context of an extended NJL-model
for mesons as quark-antiquark excitations, and calculate
the vector and pseudoscalar mesonic modes that correspond to 
the $\rho$-meson and $\pi$-meson. We then calculate the 
$\rho\pi\pi$ vertex
which is an important element  in the pion loop 
correction and also for  $\pi\pi$ scattering.
The physical process $\rho\rightarrow \pi\pi$ is then incorporated
into the $\rho$-meson self-energy as a higher order correction in
the inverse number of colors $1/N_c$ in the NJL model.
The calculations of $\pi\pi$ scattering amplitude
and cross section are outlined in Sec.~\ref{sect:pipi}.
Numerical results are presented in Sec.~\ref{sect:numer},
where we show the variations of the position of the $\rho$-pole 
with coupling strength and temperature, and the temperature 
dependence of the $\pi\pi$ cross section.
In Sec.~\ref{sect:concl}, we summarize our results and their implications.
Some calculations of quark loop contribution to the self-energy of different
mesonic modes are collected in Appendix~\ref{sect:loop}.

\section{Mesons in the extended Nambu-Jona-Lasinio model}
\label{sect:NJL}
%\subsection{Lagrangian and mesons}
Our starting point is the extended two-flavor Nambu-Jona-Lasinio Lagrangian
density:
\begin{eqnarray}
{\cal L} = & \bar\psi & (i\partial\! \! \! /-m_{0})\psi 
       + G_1\left[(\bar\psi\psi)^2 +
                          (\bar\psi i\gamma_5\tau^a\psi)^2\right]
         \nonumber \\ 
     &-& G_2\left[(\bar\psi\gamma^\mu\tau^a\psi)^2 +
                          (\bar\psi \gamma^\mu\gamma_5\tau^a\psi)^2\right],
\label{lag}
\end{eqnarray}
with the spinors $\psi$
in Dirac, color and flavor space, 
and $m_0$ a small common current mass 
for the  up and down quarks, $m_0^u=m_0^d=m_0$. The 
 $\tau^a$, $ a=1,2,3$, represent the isospin Pauli 
matrices, and $G_1$ and  $G_2$ are coupling constants of dimension [Mass]$^{-2}$. 
Eq.~(\ref{lag})
 is an extended version of the original NJL model which also includes
terms associated with the 
vector and axial-vector meson modes. 
In the chiral limit, 
$m_0=0$, this Lagrangian displays $SU_L(2)\otimes SU_R(2)$
chiral symmetry 
even for different coupling constants 
$G_1$ and $G_2$ in the scalar-pseudoscalar and vector-axial-vector sectors.
The NJL model is not a renormalizable field theory
due to the fact that the interaction between quarks is 
assumed to be point-like. It is therefore necessary to specify a 
regularization scheme in order to define the NJL model completely. 
Throughout this work, we shall employ a three-momentum cutoff
scheme in which we introduce a parameter $\Lambda$, such that
$p<\Lambda$. This cutoff $\Lambda$, 
together with $m_0$, $G_1$ and $G_2$, are the four parameters in our present 
model. 
Note that this Lagrangian has already been studied in detail elsewhere
\cite{sKli90,lemmer,mTak91}. 
We follow the notation of Ref.~\cite{lemmer} and refer the reader
to this paper for extensive detail.
Therefore 
in the following subsections, we only present
those details which  we directly require for the calculations of the 
$\pi\pi$ scattering in the $\rho$-meson channel.

\subsection{Mesons as quark-antiquark excitations}
\label{sect:meson}
Mesons as $q\bar q $ bound states or resonances can be 
obtained from the quark-antiquark scattering amplitude. For example,
in the present model,
there are collective modes carrying the quantum numbers $J^{PC}=0^{-+}$, 
$0^{++}$, $1^{--}$, and $1^{++}$
 that are composed of interacting $q\bar q$ pairs,  and which have 
the same quantum numbers as the $\pi$, $\sigma$, $\rho$, and $a_1$ mesons. 
For brevity, we shall refer to these modes as meson states of the corresponding
name in the following discussions. 

For our purpose of studying $\pi\pi$ scattering, we require
the effective 
propagators of the $\pi$ and $\rho$-mesons, as well as the effective vertex
of the $\pi q  \bar q$ interaction, all of which can be found  by 
constructing the quark-antiquark scattering amplitude.
We list these results here first.

\begin{figure}
\epsfig{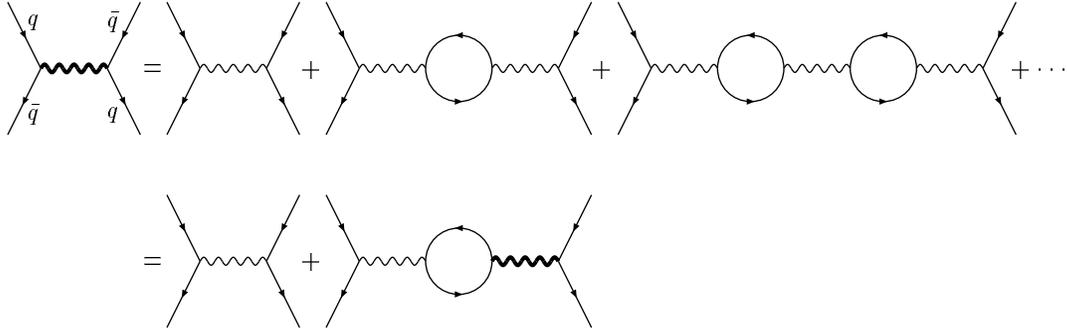}
\bigskip
\caption[]{Quark-antiquark scattering in the random phase approximation,
and its recast in the context of the Schwinger-Dyson equation.
Thick wavy lines indicate the full propagator of $\rho$-meson, and thin 
wavy lines the bare propagator.}
\label{fig:qqbar}
\end{figure}

Consider first the vector meson exchange channel, since in the present
model,
this channel is simplest as it 
does not couple with other channels. 
In the random phase approximation, the quark-antiquark 
scattering amplitude 
shown in Fig.~\ref{fig:qqbar} can be expressed to leading order in 
$1/N_c$ as an infinite sum of quark-loop chains, and can also be recast into 
the  form of a Schwinger-Dyson equation. The $\rho$-meson exchange amplitude
$D_{ab}^{\mu\nu}(q^2)$  for
quark-antiquark scattering can be generated by the one-loop polarization
function $\Pi^{VV}_{\mu\nu,ab}(q^2)$ via  the Schwinger-Dyson equation 
 of Fig.~\ref{fig:qqbar},
\be
\left[ -iD^{\mu\nu}_{ab}\right]
= \left[ -2iG_2\delta_{ab} g^{\mu\nu} \right]
+\left[ -2iG_2\delta_{ac} g^{\mu\lambda} \right]
\left[-i\Pi^{VV}_{\lambda\sigma,cd} 
\right] \left[-iD^{\sigma\nu}_{db} \right],
\label{e:rhoSD}
\ee
where $a,b,c,d$ are isospin indices, and $\mu$, $\nu$ Lorentz indices. 
The notation $VV$ refers to the vector-vector channel. 
The calculation  of the one-loop polarization function 
$\Pi^{VV}_{\mu\nu,ab}(q^2)$
is straightforward
\cite{lemmer}, and is collected for completeness
in the Appendix together with some other one-loop
polarization functions for different channels that we need in the 
calculations below.
The Lorentz and flavor structure 
for the $\rho$-meson channel allows for 
the decomposition into transverse ($T$) and longitudinal ($L$) 
components, \ie
\begin{eqnarray}
&\Pi^{VV}_{\mu\nu,ab}(q^2)&=
\left[\Pi^{VV}_T(q^2)T_{\mu\nu}+\Pi^{VV}_L(q^2)L_{\mu\nu}
\right]\delta_{ab},
\label{e:VV}
\end{eqnarray}
where
$L^{\mu\nu}=\hat{q}^\mu\hat{q}^\nu$ and $T^{\mu\nu}=g^{\mu\nu}-
\hat{q}^\mu\hat{q}^\nu$ with $\hat{q}^\mu=q^\mu/\sqrt{q^2}$.
Consequently 
the $\rho$-meson exchange amplitude can also
be decomposed into transverse and 
longitudinal components as
\be
D^{\mu \nu}_{ab}(q^2)&=&
\left[ D_T(q^2) T^{\mu \nu}
+D_L(q^2) L^{\mu \nu}\right] \delta_{ab}, 
\ee
% are
%the standard longitudinal and transverse tensors,
%and $q$ is the four-momentum of the meson mode.
where 
\be
-iD_{T,L}(q^2) &=& {-2iG_2 \over 1+2G_2\Pi^{VV}_{T,L}(q^2)}.
\label{e:dtrho}
\ee
It will be seen later that only the transverse exchange amplitude
 of the $\rho$-meson
enters into the $\pi\pi$ scattering amplitude.

In the pion channel, the situation is more complicated because of 
the mixing between the pseudoscalar ($\pi$) and axial-vector 
($a_1$) channels. In analogy  to the case in the $\rho$-meson channel, the
Lorentz and flavor structure of the relevant one-loop polarizations
\cite{lemmer} (see
the Appendix) leads to the following decompositions,
%\begin{mathletters}
\begin{eqnarray}
&D^{AA}_{\mu\nu,ab}(q^2) &=[D^{AA}_T(q^2)
T_{\mu\nu}+D^{AA}_L(q^2) L_{\mu\nu}]\delta_{ab}
\label{e:AA}
\\
&D^{AP}_{\mu,ab}(q^2) &=D^{AP}(q^2) \hat{q}_\mu \delta_{ab},
\label{e:AP}
\\
&D^{PA}_{\mu,ab}(q^2) &=D^{PA}(q^2) \hat{q}_\mu \delta_{ab},
\label{e:PA}
\\
&D^{PP}_{ab}(q^2)&=D^{PP}(q^2)\delta_{ab},
\label{e:PP}
\end{eqnarray}
%\end{mathletters}
where the indices $A$ and $P$ refer to the axial and pseudoscalar channels,
respectively.
The Schwinger-Dyson equations 
become a set of coupled 
equations that can be written in matrix form as
%
%\begin{mathletters}
%\begin{eqnarray}
%-iD^{PP}&=& 2iG_1
%+2iG_1\left[ -i\Pi^{PP}\right] \left[-iD^{PP}\right]
%+2iG_1\left[-i\Pi^{PA}\right] \left[-iD^{AP}\right]
%\label{e:PPS}
%\\
%-iD^{PA}&=& 2iG_1\left[ -i\Pi^{PP}\right] \left[-iD^{PA}\right]
%+2iG_1\left[-i\Pi^{PA}\right] \left[-iD^{AA}_L\right]
%\label{e:PAS}
%\\
%-iD^{AP}&=& -2iG_2\left[ -i\Pi^{AP}\right] \left[-iD^{PP}\right]
%-2iG_2\left[-i\Pi^{AA}_L\right] \left[-iD^{AP}\right]
%\label{e:APS}
%\\
%-iD^{AA}&=& -2iG_2 -2iG_2\left[ -i\Pi^{AP}\right] \left[-iD^{PA}\right]
%-2iG_2\left[-i\Pi^{AA}_L\right] \left[-iD^{AA}_L\right]
%\label{e:AAS},
%\end{eqnarray}
%\end{mathletters}
%Note that the transverse axial-vector polarization $\tilde\Pi^{AA}_T(q^2)$
%decouples from other channels.
%If we define two $2\times 2$ matrices as,
\begin{eqnarray}
-i {\bf D}=i{\bf K}+i {\bf K} \left[-i{\bf \Pi} \right] \left[-i{\bf D}
\right],
\end{eqnarray}
where
\begin{eqnarray}
-i\bf{D}=
\left(\begin{array}{cc}
-iD^{PP}&-iD^{PA}\\
-iD^{AP}&-iD^{AA}_L
\end{array}\right)\;,
\qquad
\bf{K}=
\left(\begin{array}{cc}
2G_1 & 0 \\
0 & -2G_2
\end{array}\right)\; .
\label{e:mtrx}
\end{eqnarray}
This matrix equation has the solution
\begin{eqnarray}
-i{\bf D}&=&
i{\bf K}(\bf{ 1-\Pi\,K})^{-1}
\nonumber \\
&=& {2iG_1 \over {\cal D}}
\left(\begin{array}{cc}
A & iB \\
-iB & C
\end{array}\right),
\end{eqnarray}
where $A=1+2G_2\Pi^{AA}_L$, $B=2iG_2\Pi^{PA}$,
$C=-G_2(1-2G_1\Pi^{PP})/G_1$,  and 
${\cal D}={\rm Det}({\bf 1-\Pi K})$.
%=
%(1-2G_1\Pi^{PP})(1+2G_2\Pi^{AA}_L)+4G_1G_2\Pi^{PA}\Pi^{AP}$.

In order to resolve the mixing between pseudoscalar and axial-vector modes
and identify the corresponding independent eigenstates and their coupling
to the quarks, 
the matrix of the coupled exchange amplitudes
which appears in the
quark-antiquark scattering amplitude,
\begin{eqnarray}
-i{\cal T}& =&
\left(i\gamma_5\tau,\;\;\;\; 
 \gamma_\mu\gamma_5\tau\right)
\left(\begin{array}{cc}
-iD^{PP} & -iD^{PA}\hat{q}_\nu \\
-iD^{AP}\hat{q}_\mu & -iD^{AA}_L L_{\mu\nu}
\end{array}\right)
\left(\begin{array}{c}
i\gamma_5\tau 
 \\
\gamma^\nu\gamma_5\tau
\end{array}\right),
%\nonumber \\
%&=& -i{\cal T}_\pi + -i{\cal T}_a.
\end{eqnarray}
has to be diagonalized.    One obtains the form
\begin{equation}
-i{\cal T}= -i{\cal T}_\pi  -i{\cal T}_a,
\end{equation}
 with
$-i{\cal T}_\pi$ and $-i{\cal T}_a$ being 
the $q\bar q$ scattering amplitudes for the 
pure pseudoscalar ($\pi$) and axial-vector ($a_1$) exchange, respectively.
For our purpose, we examine only $-i{\cal T}_\pi$ and approximate it near the pole
position as 
\be
-i{\cal T}_\pi
\sim
[(ig_{\pi qq}-ig_{pv}{\not\!\hat{q}\over 2m_q})i\gamma_5\tau]
\otimes
[(ig_{\pi qq}+ig_{pv}{\not\!\hat{q}\over 2m_q})i\gamma_5\tau]
{i\over q^2-m_{\pi}^2},
\ee
from which we can read off the complete $\pi q \bar q$ vertex
to be 
\begin{equation}
(ig_{\pi qq}-ig_{pv}{\not\!\hat{q}\over 2m_q})i\gamma_5\tau.
\end{equation}
In this expression,
 the relevant coupling constants $g_{\pi qq}$ and $g_{pv}$ 
are given as
\be
g^2_{\pi qq}=-2G_1(1+2G_2\Pi^{AA}_L)
\left.
\left({\partial {\cal D} (q^2)
\over \partial q^2} \right)^{-1}
 \right|_{q^2=m_\pi^2},
\label{e:g}
\ee
and
\be
g^2_{pv}=2G_2(1-2G_1\Pi^{PP})
\left({4m_q^2\over q^2}\right)
\left. \left({\partial {\cal D} (q^2)
\over \partial q^2} \right)^{-1}
 \right|_{q^2=m_\pi^2}.
\label{e:gv}
\ee
The additional pseudovector coupling constant $g_{pv}$ 
is induced by
the pseudoscalar-axial-vector mixing 
\cite{sKli90}.

Note that in this section, our formulae have been explicitly stated for
temperature $T=0$.    The generalization to finite temperature leaves these
equations formally unaltered - in that case, however, Lorentz invariance
is broken and the functions are no longer a function of $q^2$, rather 
of $q_0^2$ and $\vec q^2$ separately.   The derivatives required in 
Eqs.(\ref{e:g}) and (\ref{e:gv}) are then to be taken with respect to
$q_0^2$ and evaluated at $q_0^2=m_\pi^2$, and ${\bf q}^2=0$.
However, in the following,
for brevity we will still use 4-momentum argument in the expressions
even for finite temperature, and they should be understood
in the above sense.
  
\subsection{$\rho \pi \pi$ vertex}
\begin{figure}
\epsfig{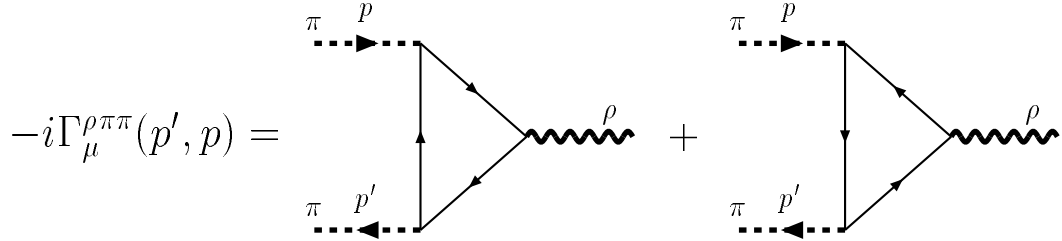}
\bigskip
\caption[]{$\rho\pi\pi$ vertices to the lowest order in $1/N_c$. the two diagrams
differ by the direction in the quark loop. They contribute equally as explained in
the text. 
\label{fig:rhopipi}}
\end{figure}

 The $\rho \pi \pi$ vertex  is an essential
 element which is required for constructing the pion loop graph and 
 the $\pi\pi$ scattering amplitude.
As will be seen in the next subsection,
the $\rho$-meson may couple to pions via a triangular quark loop which
is shown in Fig.~\ref{fig:rhopipi}.
Using Feymann rules for finite temperature, one may write
\begin{eqnarray}
-i\Gamma^{\rho \pi\pi}_\mu (p^\prime,p)=
2\times (-1)\times
%\int {d^4k \over (2\pi)^4} 
{i\over \beta} \sum_n \int^\Lambda {d^3k 
\over (2\pi)^3}
{\rm Tr}[
(ig_{\pi qq}+ig_{pv}{\not\!\hat{p}\over 2m_q})i\gamma_5
iS(k)
\nonumber \\
\times
(ig_{\pi qq}-ig_{pv}{\not\!\hat{p^\prime}\over 2m_q})i\gamma_5
iS(p^\prime +k)\gamma_\mu iS(p+k) ]\; ,
\label{e:gamma}
\end{eqnarray}
where the trace Tr[$\cdots$] is to be
carried out over the color and spinor indices,
$\beta$ is the inverse temperature,
and the Matsubara sum in $n$ runs over the fermionic frequencies
$\omega_n=(2n+1)\pi/\beta$, with $n=0, \pm 1, \pm 2, \cdots$,
which via $k=(i\omega_n,{\bf k})$
enter into the quark propagator $S(k)$.
%=(\not\! k +m_q)/(k^2-m_q^2)=
%[(i\omega_n)\gamma_0-{\bf k\cdot \gamma}+m_q)/[(i\omega_n)^2-E_k^2]$,
%with $E_k^2={\bf k}^2+m_q^2$.
%After the Matsubara summation the limit $\eta \rightarrow 0$ is to be taken,
%and an analytic continuation of the momenta to the real variables is to be
%done.
%applied to the momenta 
%$q=(i\mu_m,{\bf q})$, $\mu_m=2m\pi/\beta$,
%$m=0, \pm 1, \pm 2, \cdots$, which implies $i\mu_m\rightarrow
%q_0+i\epsilon$, with $\epsilon\rightarrow 0^+$.
Note that the $\rho q \bar q$ vertex is represented in Eq.~(\ref{e:gamma})
only in its spinor structure $\gamma_\mu$. The coupling strength 
$g_{\rho q\bar q}$ will always be included via the $\rho$-meson exchange amplitude
into the scattering amplitude, which is our convention.
In deriving Eq.~(\ref{e:gamma}), we point out that both graphs of 
Fig.~\ref{fig:rhopipi} give equal contributions, since they differ from
each other for the Dirac trace of the quark loop only by a minus sign, which
will be cancelled out by another minus sign arising from the flavor factors.
This is the origin of the factor 2 in Eq.~(\ref{e:gamma}).
Note that the flavor factors have been omitted in Eq.~(\ref{e:gamma}),
and will be incorporated later.
We have also introduced the three-momentum cut-off $\Lambda$ to regularize
the divergence of the integral.

Direct evaluation of Eq.~(\ref{e:gamma}) and an analytic 
continuation to real variables leads us to the form
\be
-i\Gamma^{\rho \pi\pi}_\mu (p^\prime,p)=
-iV_{\rho\pi\pi}(p^\prime,p) (p^\prime+p)_\mu
\label{gammarpp} 
\ee
for the vertex, where the dimensionless function $V_{\rho\pi\pi}$ 
can be expressed in terms of several elementary integrals 
$I_1$, $I_2(p^\prime-p)$,
$I_3(p^\prime,p)$, whose definitions 
are given in Eqs.~(\ref{e:I1})-(\ref{e:I3}) of the Appendix. One finds
\begin{eqnarray}
-iV_{\rho\pi\pi}(p^\prime,p)= &-& 4iN_c
\left\{ g_{\pi qq}(g_{\pi qq}+g_{pv})
I_2(p) \right.
\nonumber \\
&+&  {(g_{\pi qq}+g_{pv})^2 \over q^2-4m_\pi^2}
\left[(q^2-2m_\pi^2)[I_2(p^\prime-p) -I_2(p)]
+2m_\pi^4 I_3(p^\prime,p) \right ]
\nonumber \\
&-& \left. {g_{pv}^2 \over 12m_q^2}
\left[ (q^2+2m_q^2)I_2(p^\prime-p)-2I_1 \right]
\right\},
\label{grpp} 
\end{eqnarray}
where 
the momentum of the $\rho$-meson is denoted by $q=p-p^\prime$,
and $N_c$ is the number of colors.

\subsection{$1/N_c$ corrections}
\begin{figure}
\epsfig{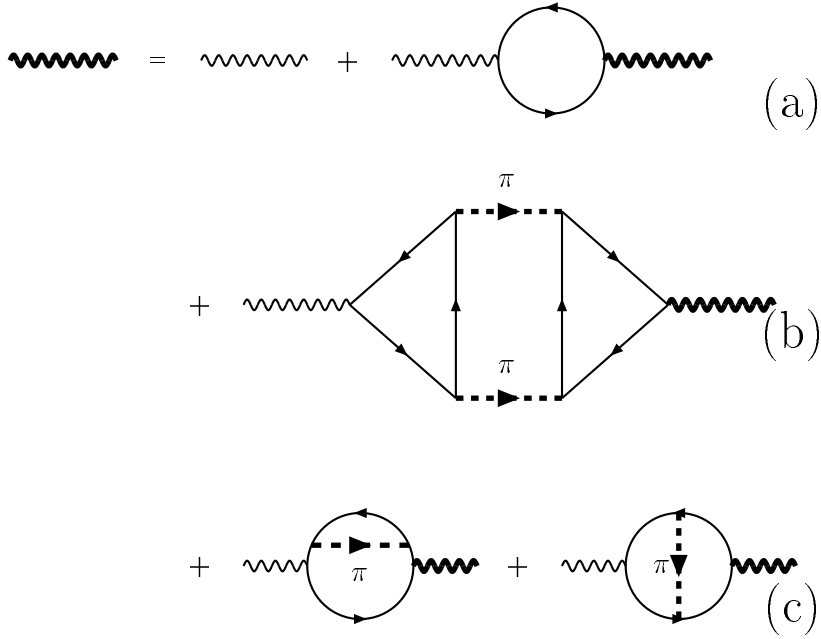}
\bigskip
\caption[]{Quark loop and pion loop contributions to the $\rho$
propagator: (a) represents the leading order contribution in $1/N_c$ 
expansion, while (b) and (c) are the next 
order corrections. Thick wavy lines indicate 
the full propagator of $\rho$-meson, and thin
wavy lines the bare propagator.}
\label{fig:rhoqpi}
\end{figure}

%\subsection{The pion loop contribution}
In this section, we extend the calculation of the $\rho$-meson (vector-vector)
 polarization 
to the next order in the $1/N_c$ expansion. 
There are two types of diagrams that enter at this level, 
and they are shown as 
(b) and (c) of Fig.~\ref{fig:rhoqpi}, 
{\it i.e.}  we construct
\begin{equation}
\Pi_{\mu\nu}^{{\rm total}} = \Pi_{\mu\nu}^{(q)} + \Pi_{\mu\nu}^{(\pi)}
\end{equation}
where $\Pi_{\mu\nu}^{(\pi)}= \Pi_{\mu\nu}^{(\pi~\rm loop)}
+ \Pi_{\mu\nu}^{(\pi~\rm tad)} $, and
$\Pi_{\mu\nu}^{(\pi~\rm loop)}$
and $\Pi_{\mu\nu}^{(\pi~\rm tad)}$ refer to the contributions of 
Fig.~\ref{fig:rhoqpi} (b) and (c), respectively. 
   The function $\Pi_{\mu\nu}^{(q)}$
refers
to the vector-vector polarization of a single quark loop as already
introduced
in Eq.(\ref{e:VV}), $\Pi_{\mu\nu}^{(q)} = \Pi_{\mu\nu}^{VV}$, and which is
given in the Appendix, see Eqs.(\ref{e:a7})
 and (\ref{e:a8}). 
The additional term $\Pi_{\mu\nu}^{(\pi~\rm loop)}$ that contains the pion loop
 is necessary on
physical grounds, since the decay $\rho\rightarrow\pi\pi$ is the main decay channel.
Within the context of the NJL model, 
  counting arguments in the expansion in  powers of $1/N_c$ allows one to
classify this term as being of the next order in $1/N_c$ relative to the
quark bubble.   
One notes that in addition, the contribution $\Pi_{\mu\nu}^{(\pi~\rm tad)}$
is also of the same order. It is required on physical grounds to ensure 
current conservation \cite{lem95}.
Note that if one were to contract the quark 
loops of Fig.~\ref{fig:rhoqpi}(c) to
a point, one would recover the tadpole diagrams of Ref.~\cite{gal91}.
We will also see in Sec.~\ref{sect:pipi}
 that this choice of diagrams 
enables us to guarantee unitarity of the $\pi\pi$ scattering 
amplitude.
    With the introduction of this extra contribution,
 the quark-antiquark scattering amplitude of Eq.(\ref{e:dtrho}) 
becomes
modified to read
\be
-iD_{T,L}(q^2)& =&
{-2iG_2 \over 1+2G_2\left[
\Pi_{T,L}^{(q)}(q^2)+\Pi_{T,L}^{(\pi)}(q^2)\right]}.
\label{rhotrans}
\ee

In the pole approximation to the pion propagator,
the $\rho$ self-energy arising from the pion loop as shown in 
Fig.~\ref{fig:rhoqpi}(b),
is given by:
\begin{eqnarray}
-i\Pi^{(\pi~\rm loop)}_{\mu\nu}(q)
={i\over \beta} \sum_n e^{i\mu_n \eta}
\int^\Lambda {d^3p \over (2\pi)^3}
 [&-&i\Gamma_\mu^{\rho\pi\pi}(q-p,-p)f_{\rho\pi\pi}]
\times
[-i\Gamma_\nu^{\rho\pi\pi}(p,p-q)f_{\rho\pi\pi}]
\nonumber \\
&&\times {i\over p^2-m_\pi^2}
\times {i\over (q-p)^2-m_\pi^2}. \nonumber \\
&&
\label{pi2}
\end{eqnarray}
Here the pion momentum $p$ is in the imaginary time formalism
$p=(i\mu_n,{\bf p})$, with the bosonic Matsubara frequencies
$\mu_n=2n\pi /\beta$, $n=0, \pm 1, \pm 2, \cdots$.
After evaluation of the Matsubara sum over $n$, the external $\rho$
momentum $q=(i\nu_n,{\bf q})$ is to be analytically continued to   
construct a causal propagator, $m_\pi^2\rightarrow m_\pi^2-i\epsilon$.
Due to the complicated structure of the $\rho \pi \pi$ vertex
given in Eq.~(\ref{gammarpp}), however,
the complete calculation of the real part of $\Pi^{(\pi~\rm loop)}$
is quite difficult.
%As we shall see in Sec.~\ref{analyt},
We shall instead approximate in Eq.~(\ref{pi2}) the complete $
[-i\Gamma_\mu(
p^\prime,p)]$ by $[-iV_{\rho\pi\pi}(\tilde p,\tilde p)](p+p^\prime)_\mu$
with $\tilde p=(q_0/2,\sqrt{q_0^2/4-m_\pi^2})$,
and $q_0$ the zeroth component of the 
$\rho$ momentum.
%with all pion
%momenta inside $[-iV_{\rho\pi\pi}(q^2)]$ on-shell,
The $\rho \pi \pi$ vertex is then independent
of the integration variable and can be factored out of the integration.
%Inserting the result for the $\rho\pi\pi$ vertex given in
%Eq.~(\ref{gammarpp}),
We find the transverse and longitudinal parts of the
pion loop contribution to have  the form   
\begin{eqnarray}
\Pi^{(\pi~\rm loop)}_T(q^2)
&=& -{i\over 3}[-iV_{\rho\pi\pi}(\tilde p,\tilde p)f_{\rho\pi\pi}]^2
 \left[ (-q^2+4m_\pi^2)I_2^{(\pi)}(q)
                          +2I_1^{(\pi)} \right],
\label{e:pitpi}
\\
\Pi^{(\pi~\rm loop)}_L(q^2)
&=& -2i[-iV_{\rho\pi\pi}(\tilde p,\tilde p)f_{\rho\pi\pi}]^2
         I_1^{(\pi)},
\end{eqnarray}
where the integrals $I_1^{(\pi)}$ and $I_2^{(\pi)}(p)$ have
 the same structure as $I_1$ and $I_2(p)$
given in Eqs.~(\ref{e:I1}) and (\ref{e:I2}) in the Appendix, except that
the quark mass $m$ is to  be replaced by the pion mass $m_\pi$,
and the fermionic frequencies $\omega_n=(2n+1)\pi/\beta$ are to be 
replaced by the 
bosonic frequencies $\mu_n=2n\pi/\beta$ with $n=0, \pm 1, \pm 2, \cdots$.
An analytic expression for the 
 imaginary part of $\Pi^{(\pi \rm loop)}_T(q^2)$ can also be derived.
One finds
\be
{\rm Im}\Pi^{(\pi~\rm loop)}_T(q^2)
= - [-iV_{\rho\pi\pi}(\tilde p,\tilde p)f_{\rho\pi\pi}]^2
{1\over 48\pi q_0}
(q_0^2-4m_\pi^2)^{3/2} \coth (\beta q_0/4),
\label{piloopimag}
\ee
where $q$ is the momentum of the $\rho$-meson, which can be off-shell.

One notices immediately that $\Pi^{(\pi~\rm loop)}_L(q^2)\neq 0$,
as was the case for the quark loop contribution (see Eq.~\ref{e:a8}).
A direct calculation of Figs.~\ref{fig:rhoqpi}(c) for 
$\Pi^{(\pi~\rm tad)}_{\mu\nu}$ would ensure that 
$\Pi^{(\pi)}_L=0$ \cite{lem95}. We do not do this explicitly here, but
evaluate this quantity indirectly assuming the constraint 
 $\Pi^{(\pi~\rm loop)}_L(q^2)\neq 0$.
Since $\Pi^{(\pi~\rm tad)}_{\mu\nu}=\Pi^{(\pi~\rm tad)}g_{\mu\nu}$,
it follows that $\Pi^{(\pi~\rm tad)}=-\Pi^{(\pi~\rm loop)}_L$. Thus all 
polarization functions are known.
 
\subsection{$\rho$-meson mass and analytic structure of the scattering
amplitude}
\label{sect:anal}
In order to identify the mass of the $\rho$-meson, 
one is required to find the poles of the vector-vector scattering amplitude
$D^{\mu\nu}$ or the corresponding propagator for this channel in the complex energy 
plane. 
This translates to finding the poles of the transverse
 amplitude
%\footnote{In the absence of the pion loop contribution
%$\Pi^{(\pi)}_{\mu\nu}$, one has
%\begin{equation}
%D_T = \frac{2G_2}{1+2G_2\Pi_T^{(q)}},\quad \quad  {\rm while}
%\quad\quad D_L=2G_2,
%\end{equation}
%indicating that a pole in $D^{\mu\nu}$ can arise only from the transverse
%part.   This remains true for the inclusion of $\Pi_{\mu\nu}^{(\pi)}$. 
%This point of view is confirmed in Sec.~\ref{sect:pipi} in the discussion of 
%$\pi\pi$ scattering, where one sees explicitly that 
%only the transverse part of the propagator enters
%into the scattering amplitude.}
$D_T(q^2)$ that is given in   
 Eq.(\ref{rhotrans}).
We thus investigate the analytic structure of $D_T$ from
Eq.(\ref{rhotrans}) in this section.

To avoid the complexity associated with the analytical continuation of
the $\rho\pi\pi$ vertex in the complex plane,
we shall assume a constant value for the dimensionless
$\rho\pi\pi$ vertex function $V_{\rho\pi\pi}$
in the following study of the analytic structure,
which is a simplification that will not change the feature of our results.
 
The poles  of the scattering amplitude
are given by the zeros of the denominator of $D_T(q^2)$,
which is, according to Eq.~(\ref{rhotrans}),
\be
F(q)=1+2G_2\left[ \Pi^{(q)}_T(q^2)+\Pi^{(\pi)}_T(q^2)\right].
\label{e:fq}
\ee
Before we discuss the analytic structure of the function $F(q)$, it is
useful to first
examine  the quark loop contribution $\Pi^{(q)}_T(q^2)$.
The analytic structure  of $\Pi^{(q)}_T(q^2)$ is governed by the integral
$I_2(q)$,
\be
i I_2(q)\equiv
i\times \intk {1 \over [k^2-m_q^2]\;[(k+q)^2-m_q^2]}.
\label{e:sumI2}
\ee
In  back-to-back kinematics,
$i I_2$  depends only on the zeroth component of the momentum $q$.
After evaluating of the Matsubara sum and performing
an analytical continuation,
% $q=(i\mu_n,{\bf 0}) \rightarrow (q_0+i\epsilon,{\bf 0})$,
one finds the integral in
Eq.~(\ref{e:sumI2}) to be
\be
i I_2(q_0)=-{1 \over 2\pi^2}\int_{m_q}^{\Lambda_E} dE
{\sqrt{E^2-m_q^2} \tanh (\beta E/2) \over 4E^2-q_0^2-i\epsilon}\, ,
\label{e:iIt2}
\ee
where $\Lambda_E=\sqrt{\Lambda^2+m_q^2}$.

We consider $i I_2(q_0)$ as a complex function in the complex 
$q_0$-plane.
This function has branch points at $q_0=\pm 2m_q, \pm 2\Lambda_E$, but for
our
purpose only the branch point at $q_0=2m_q$ will be relevant.
Choosing in the complex $q_0$-plane a cut starting from the branch point
 $q_0=2m_q$
and running to the right, one
has two regions on the first Riemann  sheet  in
which $i I_2(q_0)$ is to be evaluated separately:
 
(i) On the first sheet of the complex plane away from the cut,
where the integrand in Eq.~(\ref{e:iIt2})
is non-singular,
$i I_2(q_0)$
 is readily calculated as
\be
i I_2(q_0)=-{1 \over 2\pi^2}\int_m^{\Lambda_E} dE
{\sqrt{E^2-m_q^2} \tanh (\beta E/2) \over 4E^2-q_0^2}\, ,
\ee
which can be expressed analytically at zero temperature  by
\be
i I_2(q_0)\left. \right|_{T=0}
=- {1 \over 8\pi^2}
\left[ \ln {\Lambda_E +\Lambda \over m_q}
-{\sqrt{4m_q^2-q_0^2} \over q_0}
\arctan \left({q_0\Lambda \over \Lambda_E\sqrt{4m_q^2-q_0^2}}\right)
\right].
\ee
 
(ii) On the real axis for the values of $q_0$ with
$2m_q< {\rm Re}(q_0) < 2\Lambda_E$,
the integrand of $i I_2(q_0)$ develops a singularity at
$E=q_0/2$ if $\epsilon \rightarrow 0^+$. Making use of the formula
\be
\lim_{\epsilon \rightarrow 0^+} {1 \over y-i\epsilon}
={\cal P} {1 \over y} + i\pi \delta (y)
\ee
one obtains
\be
i I_2(x)=-{1 \over 2\pi^2}
{\cal P} \int_m^{\Lambda_E} dE
{\sqrt{E^2-m_q^2} \tanh (\beta E/2) \over 4E^2-x^2}
-{i \over 16 \pi} {\sqrt{x^2-4m_q^2} \over x}
\, ,
\label{e:regionI}
\ee
where ${\cal P}$ stands for the Cauchy principal value, and $x$ the real
part of
$q_0$.
At zero temperature the integral in Eq.~(\ref{e:regionI}) can be
analytically
evaluated as,
\be
i I_2(x)\left. \right|_{T=0}
= - {1 \over 16\pi^2}
%= &-& {1 \over 16\pi^2}
\left[ 2\ln {\Lambda_E +\Lambda \over m_q}
+{\sqrt{x^2-4m_q^2} \over x}\ln {x\Lambda -\Lambda_E \sqrt{x^2-4m_q^2}
                         \over x\Lambda +\Lambda_E \sqrt{x^2-4m_q^2}}
\right]
\nonumber \\
-
%&-&
{i \over 16 \pi} {\sqrt{x^2-4m_q^2} \over x}
\, .
\label{e:regionIT0}
\ee
In order to study poles
of the scattering amplitude in the complex plane which are located
close to the physical sheet,
one has to continue the analytic functions
onto the unphysical sheet. The analytic continuation
of $i I_2(q_0)$ 
through the cut into the second sheet is obtained by \cite{hue96}
\be
i I_2(q_0)\left. \right|_{\rm 2nd}
=i I_2(q_0)\left. \right|_{\rm 1st}
-{1\over 8\pi} {\sqrt{4m_q^2-q_0^2}\over q_0} \tanh \left({\beta q_0\over 4}
\right),
\ee
where ``1st'' and ``2nd'' denote the first and the second sheet. This
analytic continuation is sufficient for our investigation since we are
only interested in the poles on the unphysical sheet that is close to the
positive real axis of the first sheet.

The integral $I_2^{(\pi)}$ occurring in the pion loop contribution
$\Pi_T^{(\pi)}
(q^2)$ (see Eq.~(\ref{e:pitpi})) has the same analytic structure as
the quark loop integral $I_2$ discussed above, except that the quark mass
$m_q$ and the hyperbolic tangent  in
$I_2$ are replaced in $I_2^{(\pi)}$
by the pion mass $m_\pi$ and the hyperbolic
cotangent, respectively.
 
\begin{figure}
\epsfig{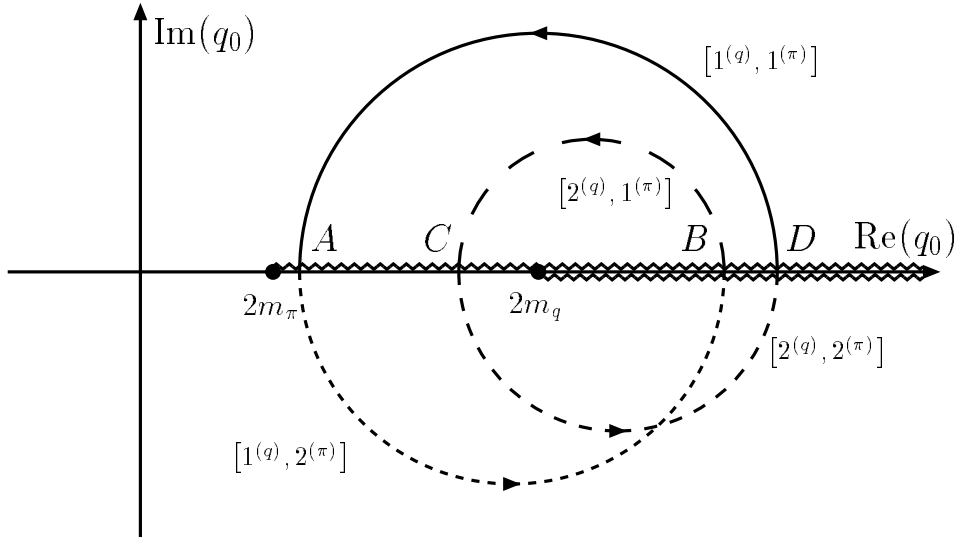}
\bigskip
\caption[]{Joining of Riemann sheets of $F(q_0)$ along two branch
cuts on the real axis 
starting from thresholds $2m_\pi$, $2m_q$ and running to the
right. Different sheets are denoted by 
$\left[ i^{(q)}, j^{(\pi)}\right]$ with $i,j=1,2$ representing 
the first, second sheet in $q\bar q$ and $\pi\pi$ channels, respectively.
(see also text.)}
\label{fig:fqsheets}
\end{figure}

The analytic structure of the function
$F(q)$ in Eq.~(\ref{e:fq}), whose zeros correspond to poles of the
scattering
amplitude, is
determined by the integrals
$I_2$ in $\Pi^{(q)}_T(q^2)$ and $I_2^{(\pi)}$ in $\Pi_T^{(\pi)}(q^2)$
which are both two-sheeted functions.
%\footnote{The non-relativistic
%scattering amplitude has two sheets.  The relativistic amplitude
%here however has a logarithmic structure and therefore infinitely many
%Riemann sheets.}. 
Consequently
$F(q)$, as a
combination of $\Pi^{(q)}_T$ and $\Pi_T^{(\pi)}$, has two branches into
multi-sheeted
functions. In Fig.~\ref{fig:fqsheets}, we illustrate the structure of $F(q)$
in the complex $q_0$-plane: $F(q)$ has two relevant branch points at
$q_0=2m_\pi$, and at $q_0=2m_q$, and two branch cuts on the real axis
starting from $2m_\pi$, and
$2m_q$ respectively and running to the right; A continuous path is drawn
to show the joining of Riemann sheets along the branch cuts. We use
 the notation
$\left[i^{(q)},j^{(\pi)}\right]$ with $i,j=1,2$ to denote different sheets,
 \eg,
$\left[1^{(q)},2^{(\pi)}\right]$ stands for the case that $I_2$ in 
$\Pi^{(q)}_T$
takes value on the first sheet, and $I_2^{(\pi)}$ in $\Pi_T^{(\pi)}$
on the second sheet, and so on.
In order to familiarize the reader with the analytic structure, 
Fig.~\ref{fig:fqsheets} shows a path through the multi-sheeted plane: 
starting
from point $A$ in the plane $\left[1^{(q)},1^{(\pi)}\right]$,
 crossing the cut into the
plane $\left[1^{(q)},2^{(\pi)}\right]$, and following the path 
until at $B$ one crosses
again a cut into $\left[2^{(q)},1^{(\pi)}\right]$, then turning around to $C$ 
and entering
into $\left[2^{(q)},2^{(\pi)}\right]$, \etc.

The numerical evaluation of $F$ and the ensuing determination of the pole
position of $D_T(q^2)$ for the $\rho$-meson mass, is discussed in Sec.~IV.

\section{$\pi\pi$ scattering}
\label{sect:pipi}
\subsection{$s$-channel $\rho$-meson exchange}
Within the extended NJL model, there are several types of diagrams that
contribute to $\pi\pi$ scattering.
The full complement of processes  can be found in Ref.~\cite{party}.
 Among these are the box and 
$\sigma$ meson exchange diagrams, which dominate the 
$s$-wave scattering and quantitatively reproduce the scattering lengths,
 but do not suffice to account for the full scattering amplitude
as a function  of the center of mass energy
in the vector-isovector channel \cite{Donoghue}. 

\begin{figure}
\epsfig{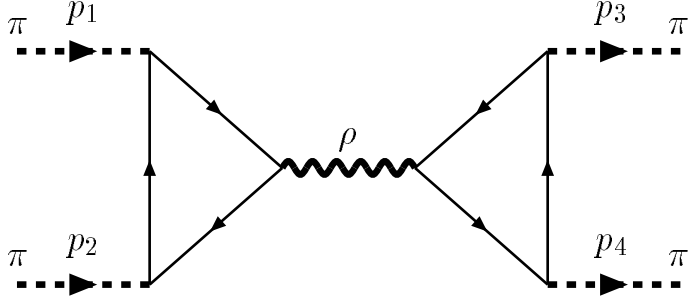}
\bigskip
\caption[]
{$\pi\pi$ scattering via $s$-channel $\rho$-meson exchange to the lowest
order in $1/N_c$.}
\label{fig:scatt}
\end{figure}

In this work, 
our interest lies in particular in  studying  the physical situation in the
mid-rapidity region in  high energy
heavy-ion collisions, where thermal
pions are copiously produced with kinetic energy, $E_\pi\sim 200$~MeV,
which is of
the order of the typical temperature scale reached
by the system shortly after the
collisions. At that stage, we expect $\pi\pi$ scattering to be
characterized by a center-of-mass energy $\sqrt{s}\geq 2E_\pi \sim 400$~MeV.
In this energy region,
according to the
phase shifts data of $\pi\pi$ scattering \cite{cdFro77},
the $\pi\pi$ cross section has already an important contribution from
the $\rho$-meson
exchange channel.
Thus, in the following, we shall  restrict ourselves to $\rho$ 
meson exchange diagrams in the $s$-channel which contribute to the $\pi\pi$
scattering at  the lowest order in 
$1/N_c$  in the extended NJL model as is shown in
Fig.~\ref{fig:scatt}.
The intermediate $\rho$-meson state in this graph is to be understood
as being given by the full propagator corresponding to Fig.~\ref{fig:rhoqpi}
and this is built up of quark-antiquark excitations as well as $\pi\pi$
modes.
 Of all  diagrams 
 we evaluate  only the $s$-channel contribution in Fig.~\ref{fig:scatt}, since 
 it should be the dominant contribution to the cross section. 
The scattering amplitude for  this graph is given as 
\be
-i{\cal M}_{\pi\pi\rightarrow \pi\pi}
= [-i\Gamma^{\rho \pi\pi}_\mu (-p_2,p_1)f_{\rho\pi\pi}]
[-iD^{\mu\nu}(p_1+p_2)]
[-i\Gamma^{\rho \pi\pi}_\mu (p_4,-p_3)f_{\rho\pi\pi}],
\label{mag} 
\ee
where $p_1$, $p_2$ and $p_3$, $p_4$ are the momenta of incoming 
and outgoing pions, respectively, 
and $f_{\rho\pi\pi}$ is the flavor
factor to be associated with
 the $\rho\pi\pi$ vertex. For example, $f_{\rho\pi\pi}=2$ for a 
$\rho^0\pi^+\pi^-$ vertex in Fig.~\ref{fig:scatt}.
At finite temperature, Lorentz 
covariance is broken, \ie, in general the cross section of 
a two-body reaction does not only depend
on the temperature, but also on the c.m. velocity of the initial pair
with respect to the heat bath.
In the following calculations we shall assume the initial pion
system to be at rest in the heat bath.

Inserting the Lorentz structure of the 
$\rho\pi\pi$ vertex given in 
Eq.~(\ref{gammarpp}),
the scattering amplitude in Eq.~(\ref{mag})
 can then be recast as
\be
%-i{\cal M}=-i [-iV_{\rho\pi\pi}(-p_2,p_1)f_{\rho\pi\pi}]^2
-i{\cal M}_{\pi\pi\rightarrow \pi\pi}
=-i [-iV_{\rho\pi\pi}(-p_2,p_1)f_{\rho\pi\pi}]^2
D_T(s)(t-u),
\label{e:ppmag}
\ee
where $s$, $t$, and $u$ are the 
standard Mandelstam variables, $s=(p_1+p_2)^2$,
$t=(p_1-p_3)^2$, and $u=(p_1-p_4)^2$. Note that
the scattering amplitude is proportional to $(t-u)=(s-4m_\pi^2)\cos\theta
=(s-4m_\pi^2)P_1(\cos \theta)$, which is consistent with the
expectation that
$\rho$-exchange is a $p$-wave scattering.
Only the transverse component of the 
$\rho$-exchange amplitude contributes to the scattering
amplitude due to the specific Lorentz structure of the
$\rho\pi\pi$ vertex
and the fact that pions in the initial- and final-state are all on the
mass shell.
This is in accordance with the fact that the poles of the transverse part
of the amplitude $D_T$ determine the position of the $\rho$ mass.

The differential cross section follows as
\be
{d\sigma(s,t) \over dt}=
{1 \over 16\pi s (s-4m_\pi^2)}
\left| {\cal M}_{\pi\pi\rightarrow \pi\pi} \right|^2,
\ee
and then the total cross section by
\be
\sigma (s) =\int_{t_1}^{t_0} dt
{d\sigma(s,t) \over dt}
\left[1+f_B(\sqrt{s}/2) \right]^2,
\ee
where $t_0=0$, $t_1=4m_\pi^2-s$,
$f_B(x)=1/[\exp (\beta x)-1]$, and the factor $\left[1+f_B(\sqrt{s}/2)
 \right]^2$
has been
 included in order to incorporate the Bose-enhancement effects in the
final-state phase space.

\subsection{Unitarity of the $\rho$-meson exchange amplitude}
\label{sect:unitarity}
Unitarity at zero temperature provides a strong constraint on any scattering
calculations, while the situation for finite temperature is less
clear.
It is thus important to confirm that the choice of diagrams evaluated for
the $\pi\pi$ scattering amplitude lead to a unitary $S-$matrix at $T=0$.
The $\pi\pi$ scattering amplitude in our approximation is to be 
calculated from  Eq.~(\ref{e:ppmag}).
We note that if $D_T(q^2)$ is restricted to contain the quark loop 
contribution 
  $\Pi^{(q)}_T$ alone, no unitarity condition can be fulfilled as   
 $\Pi_T^{(q)}$
 becomes imaginary only above the
$q\bar q$ threshold $\sqrt{s}\ge 2m_q$. This threshold is an unphysical
artifact of the NJL model, at least in the confined phase. 
Including the term
$\Pi^{(\pi)}_T$ opens
  the $\pi\pi$ threshold
for $\sqrt{s}\ge 2m_\pi$, and allows the 
optical theorem to  hold for $\pi\pi$ scattering.
We recall once again that, from a physical point of view, this channel is
essential, since  
nearly 100\% of the $\rho$-mesons decay  into pions.

It is apparent from this discussion that $\pi\pi\rightarrow\pi\pi$
scattering within this model forms part of a coupled channels system that
also involves the $q\bar q $ sector.   The scattering amplitude for 
$\pi\pi\rightarrow\pi\pi$ proceeding via $\rho$-meson exchange 
is only  {\it one} element in the 
two-channel $S$-matrix,
\be
S=
\left(\begin{array}{cc}
S_{q\bar q\rightarrow q\bar q} & S_{q\bar q\rightarrow \pi\pi} \\
S_{\pi\pi\rightarrow q\bar q} & S_{\pi\pi\rightarrow \pi\pi} 
\end{array}\right).
\ee
The other channels are all to be understood as proceeding via an
$s$-channel $\rho$-meson exchange, and can be envisaged from Fig.~5(a)
by simply constructing the relevant subprocess from this diagram.
The matrix elements for the $q\bar q\rightarrow q\bar q$ and 
$q\bar q\leftrightarrow \pi\pi$ are given as
\begin{equation}
-i{\cal M}_{q\bar q\rightarrow q\bar q}
=-i [-iV_{\rho q\bar q}f_{\rho q\bar q}]^2
D_T(s)(t-u),
\label{e:qqbar}
\end{equation}
and
\begin{equation}
-i{\cal M}_{q\bar q\rightarrow \pi\pi}
=-i [-iV_{\rho q\bar q}f_{\rho q\bar q}]
[-iV_{\rho\pi\pi}f_{\rho\pi\pi}]
D_T(s)(t-u).
\label{e:qpipi}
\end{equation}
Thus, noting
the explicit form of Eqs.(\ref{e:ppmag}), (\ref{e:qqbar}) and
(\ref{e:qpipi})
 with $D_T=2G_2/F$, and $F$ as given in
Eq.(\ref{e:fq}), one may factorize each term in the $S$ matrix as
\be
S=\left(\begin{array}{cc}
1+iA^2_{q\bar q}/F & iA_{q\bar q}A_{\pi\pi}/F \\
iA_{\pi\pi}A_{q\bar q}/F & 1+iA^2_{\pi\pi}/F 
\end{array}\right).
\ee
In this expression, 
 $A_{\pi\pi}$ follows directly from Eq.(\ref{e:ppmag}) to be
\begin{eqnarray}
A_{\pi\pi}^2 =
{G_2(s-4m_\pi^2)\over 12\pi} \sqrt{s-4m_\pi^2\over s}
\left[-iV_{\rho\pi\pi}f_{\rho\pi\pi}\right]^2,
\end{eqnarray}
while similar arguments can be given to extract $A^2_{q\bar q}$ as
\begin{eqnarray}
A_{q\bar q}^2 =
{G_2(s-4m_q^2)\over 12\pi} \sqrt{s-4m_q^2\over s}
\left[-iV_{\rho q\bar q}f_{\rho q\bar q}\right]^2.
\end{eqnarray}
%we can thus write the $S$ matrix in a factorized form,
%\be
%S_{\alpha\beta}=\delta_{\alpha\beta}
%-i{\Gamma_{\alpha}\Gamma_{\beta} \over F}
%\qquad \qquad
%\alpha,\beta=q\bar q, \pi\pi
%%\alpha,\beta=q,\pi
%\label{e:sab}
%\ee
Unitarity of the $S$ matrix, 
$SS^+ = 1$,
%\sum_\gamma S_{\alpha\gamma} S^*_{\beta\gamma} =1
%\ee
implies
\be
A_{\pi\pi}^2+A_{q\bar q}^2 =-2 {\rm Im}F
=-4G_2{\rm Im}\left[ \Pi^{(\pi)}_T\right]
-4G_2{\rm Im}\left[ \Pi^{(q)}_T\right].
\ee
Making use of the explicit result for ${\rm Im}\left[ \Pi^{(\pi)}_T\right]$
from Eqs.(\ref{piloopimag}) 
of the previous section
and similar result for ${\rm Im}\left[ \Pi^{(q)}_T\right]$, one can 
actually show that 
$A_{\pi\pi}^2=-4G_2{\rm Im}\left[ \Pi^{(\pi)}_T\right]$, and
$A_{q\bar q}^2=-4G_2{\rm Im}\left[ \Pi^{(q)}_T\right]$.
%According to the cutting rules, $\Gamma_{q\bar q}^2$ is 
%the imaginary part of the quark loop contribution,
%and $\Gamma_{\pi\pi}^2$ the imaginary part of the pion loop
%contribution.
Therefore, after including the pion loop 
contribution to the $\rho$-meson exchange amplitude,
unitarity can be guaranteed. 
 
For later use, we recall that
the elastic cross section in the $l^{th}$ partial wave scattering has
a unitarity bound:
\be
\sigma_l\leq {4\pi (2l+1) \over p_{cm}^2}\, ,
\ee
where $p_{cm}$ is the c.m. momentum.
Since the $\rho$-meson exchange corresponds to a  $p$-wave ($l=1$) scattering
process, the cross section should always be bounded by
$\sigma\leq 12\pi/p_{cm}^2$.

\section{Numerical results}
\label{sect:numer}

In this section, we present our numerical results.   We first discuss our
choice of parameters, which are set at $T=0$.     We then discuss the
numerical evaluation of the $\rho$-meson mass in the complex plane, and
the $\pi\pi$ scattering cross section for a constant value of the
$\rho \pi\pi $ vertex.   A complete calculation, including the momentum
dependence of the vertex function is presented in the 
ensuing subsection.

\subsection{Parameters}
Before discussing  our numerical results, we have to specify the 
four parameters in the present model, namely the current quark mass $m_0$,
coupling strengths $G_1$, and $G_2$, and the three-momentum cut-off $\Lambda$.
Our choice of parameters 
$m_0=3.7$~MeV, $\Lambda=650$~MeV, $G_1\Lambda^2=2.5$, 
$G_2\Lambda^2=2.3$, 
leads to a value of the pion mass as $m_\pi=140$~MeV, 
pion decay constant $f_\pi=85$~MeV,
and quark condensate $\langle\bar \psi \psi \rangle
=-(267~{\rm MeV})^3$, which  
corresponds to a quark mass 
$m_q=458.2$~MeV at temperature $T=0$. 
In particular, we have chosen the value of $G_2\Lambda^2$ in such a way
that at $T=0$,
the $\rho$-meson appears   as a $\pi\pi$ resonance at $m_\rho\simeq 768$~MeV.
The dynamically generated quark mass $m_q$ resulting from
this parameter set is deliberately chosen to be
 large, so that the $q\bar q$ threshold 
lies above the physical $\rho$-meson position and the 
 $\rho\rightarrow q\bar q$
channel is not open
at $T=0$.
The value for the condensate $\langle\bar\psi \psi\rangle$
is, however, still acceptable.

\subsection{$\rho$-poles and cross section for constant $\left|V_{\rho\pi\pi}
\right|$}
\label{sect:sigconst}
\begin{figure}
\epsfig{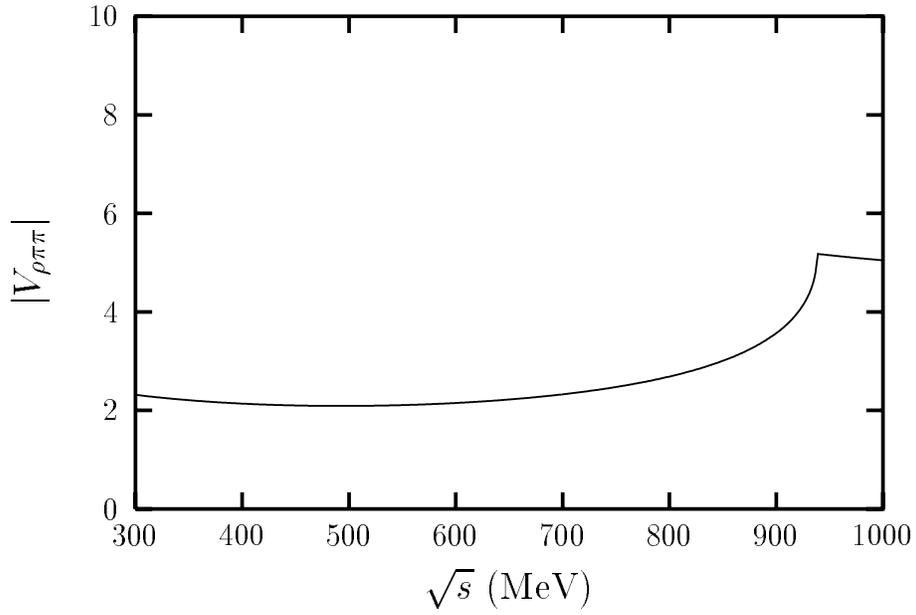}
\bigskip
\caption[]{Energy dependence of the $\rho\pi\pi$ vertex given by
Eq.~(\ref{grpp}) at temperature $T=0$. The discontinuity at 
$\sqrt{s}\simeq 920$~MeV relates to the $q\bar q$ threshold $2m_q$.}
\label{fig:vtxrpp}
\end{figure}

As in Sec.~\ref{sect:anal}, 
a constant (energy-independent) value of the $\rho\pi\pi$
vertex function
$\left|V_{\rho\pi\pi} \right|$ 
is used to avoid the intricacies induced by the complicated structure of the 
$\rho\pi\pi$ vertex.
We have checked this assumption via a direct calculation whose result
is shown in Fig.~\ref{fig:vtxrpp}.
% shows the energy dependence of the vertex given by 
%Eq.~(\ref{grpp}) at zero temperature. One sees that 
For a rather wide 
energy range below the $q\bar q$ threshold, $\left|V_{\rho\pi\pi} \right|$
depends only weakly on the energy. In this section, we will assume
 a constant value of the vertex,
$\left|V_{\rho\pi\pi} \right|=2.4$.

\begin{figure}
\epsfig{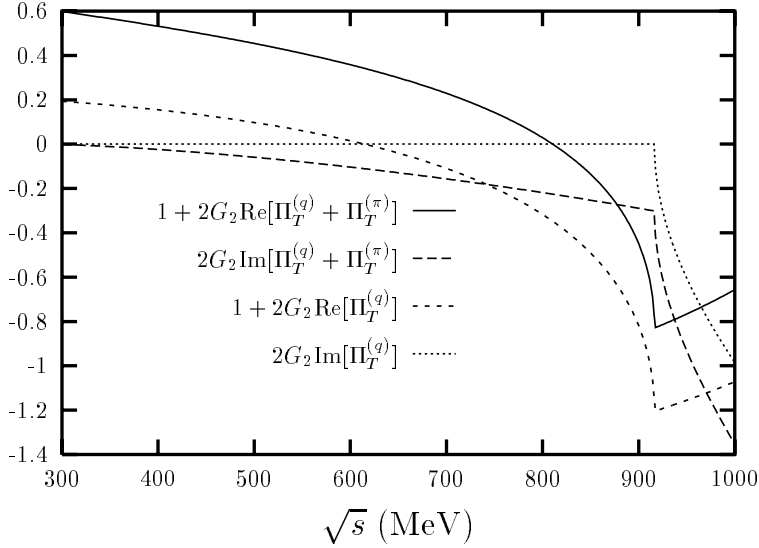}
\bigskip
\caption[]{Energy dependence of real and imaginary parts of the 
denomenator of the $\rho$-meson propagator $F$ in 
Eq.~(\ref{e:fq})
at $T=0$. The breaks at $\sqrt{s}\simeq 920$~MeV 
are due to the $q\bar q$ threshold.}
\label{fig:pivvt}
\end{figure}

Fig.~\ref{fig:pivvt} shows the energy dependence of different real
and imaginary parts of $F(\sqrt{s})$, the denominator of the transverse
$\rho$-meson exchange amplitude $D_T(s)$ given by Eq.~(\ref{e:fq}), at
the  temperature
$T=0$. From this figure,  the necessity of including
the pion loop contribution $\Pi_T^{(\pi)}$ into the $\rho$-meson self-energy
becomes apparent.
If there is no pion loop contribution to $F(\sqrt{s})$, 
below the $q\bar q$ threshold $\sqrt{s}=2m_q$, the imaginary part of the 
quark loop $\Pi^{(q)}_T$ vanishes, since the decay channel 
$\rho \rightarrow q\bar q$ is closed, and one finds a real root of 
$F(\sqrt{s})$ at $\sqrt{s}\simeq 620$~MeV where $1+2G_2{\rm Re}\left[
\Pi_T^{(q)} \right] =0$. This real $\rho$-pole
of $D_T(\sqrt{s})$ corresponds to a $\rho$-meson as a $q\bar q$ bound state,
and would lead to an unphysical divergence in the $\pi\pi$ cross section.
%at $T=0$, violating  the unitarity bound that was discussed in 
%Sec.~\ref{sect:unitarity}.

If  we take the pion loop contribution
$\Pi_T^{(\pi)}$ into account, the imaginary part of the polarization,
${\rm Im}\left [\Pi_T^{(q)}+ \Pi_T^{(\pi)} \right]
\neq 0$  below the $q\bar q$ threshold,
  as long as it is above the $\pi\pi$ threshold $\sqrt{s}=2m_\pi$.
At the position $\sqrt{s}\simeq 800$~MeV, where the real part of 
$F(\sqrt{s})$ vanishes, the imaginary part of $F(\sqrt{s})$ 
does not vanish, and we have to move into the complex energy plane to 
find a complex root $m_\rho$ of $F(q)$, which corresponds to a $\rho$-meson
as a $\pi\pi$ resonance with ${\rm Re}(m_\rho)\simeq 800$~MeV and finite width.

\begin{figure}
\epsfig{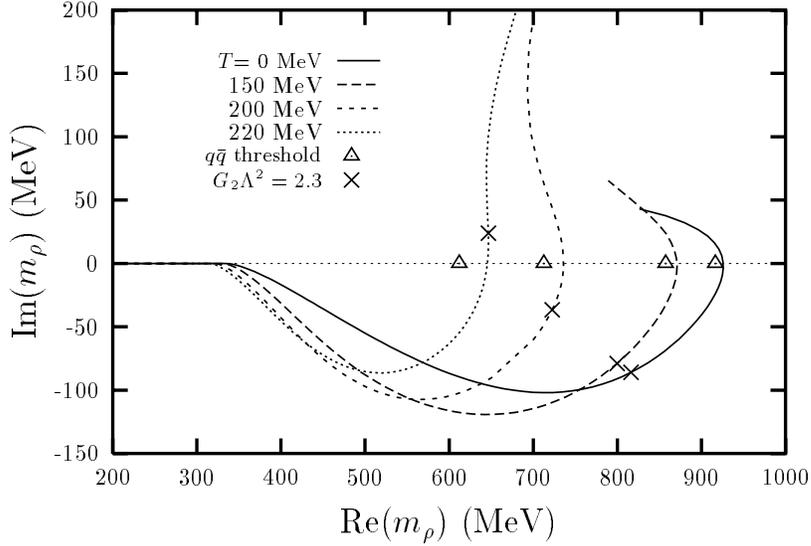}
\bigskip
\caption[b]{Variations of the $\rho$-pole position in the complex energy plane
with the coupling strength $G_2\Lambda^2$ at different temperatures. 
The trajectory at $T=0$ begins with $G_2\Lambda^2=6.4$ for
Re($m_\rho$)=200~MeV, and it moves to the right with decreasing $G_2\Lambda^2$.
Similar for other temperatures. 
For each temperature 
the $q\bar q $ threshold $2m_q$ is indicated by a 
dotted-triangle, and  
the $\rho$-pole
position for the physical parameter $G_2\Lambda^2=2.3$ is marked by a cross.}
\label{fig:rhopole}
\end{figure}

\begin{figure}
\epsfig{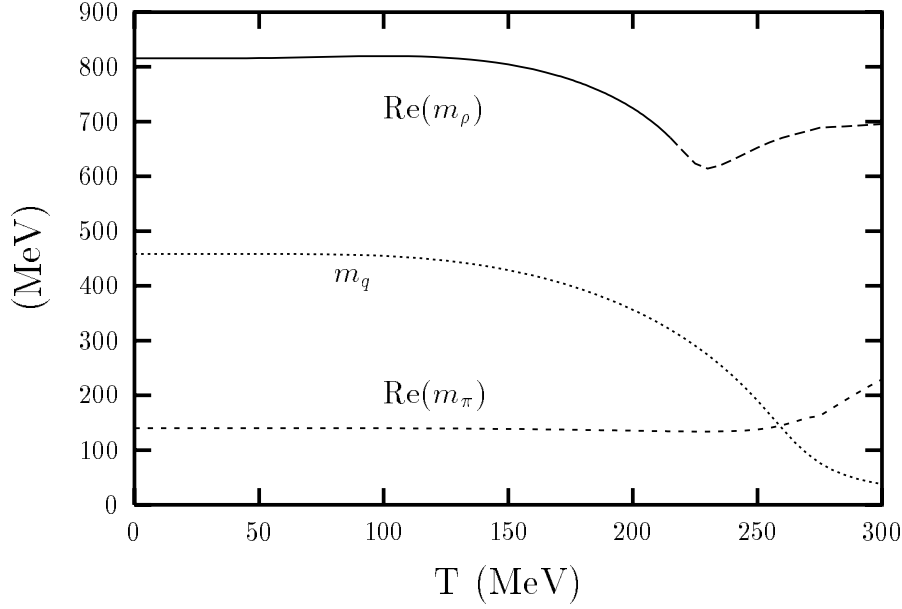}
\bigskip
\caption[]{Temperature dependence of Re($m_\rho$), Re($m_\pi$), and $m_q$.
The long-dashed line indicates the values of Re($m_\rho$) corresponding to
$\rho$-poles in the sheet $\left[2^{(q)},1^{(\pi)} \right]$.}
\label{fig:mrhot}
\end{figure}

Fig.~\ref{fig:rhopole} illustrates $\rho$-pole trajectories for different 
temperatures 
as a function of the coupling strength $G_2\Lambda^2$.
At T=0, for a very large (unphysical) value $G_2\Lambda^2=6.4$, the 
$\rho$-meson appears as a bound state at 200~MeV and vanishing imaginary
part since it is located below the $\pi\pi$ threshold. With decreasing 
values of $G_2\Lambda^2$ the $\rho$ becomes less bound, until above the 
$\pi\pi$ threshold the trajectory moves away from the real axis and 
Im($m_\rho$) becomes negative and of the order of 100~MeV.  A cross on 
the trajectory marks the physical value of $G_2\Lambda^2=2.3$. Decreasing
the value of $G_2\Lambda^2$ even more the trajectory approaches the
real axis again, and 
enters into the
sheet $\left[2^{(q)},1^{(\pi)} \right]$  
closely above the $q\bar q$ threshold.
This exercise repeated for several 
temperatures is shown in Fig.~\ref{fig:rhopole}. If one follows the position
of the crosses, which indicate the physical location of the $\rho$ as a 
function of temperature, one reads off the main result of our investigation:
very little changes for Re($m_\rho$) for temperatures between 0 and 130~MeV.
This can be more clearly seen from Fig.~\ref{fig:mrhot}, where we show the 
temperature dependence of Re($m_\rho$), Re($m_\pi$) and $m_q$. 
%The long-dashed
%line indicates the values of Re($m_\rho$) when the $\rho$-pole moves into the 
%sheet $\left[2^{(q)},1^{(\pi)} \right]$.
The pion Mott temperature $T_M$, which is defined by
$m_\pi(T_M)=2m_q(T_M)$, is found to be $T_M\simeq 280$~MeV. This somewhat 
large $T_M$ is a consequence of the large $m_q\simeq 460$~MeV at $T=0$.

\begin{figure}
\epsfig{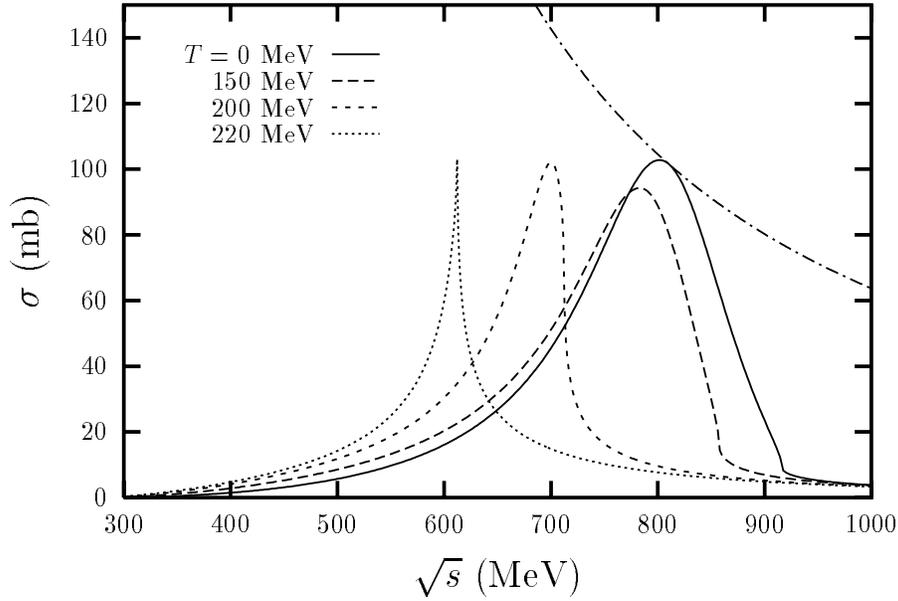}
\bigskip
\caption[]{Calculated $\pi\pi$ cross sections in the vector-isovector channel
at different temperatures for 
a constant vertex $\left|V_{\rho\pi\pi} \right|=2.4$. The unitary bound 
$12\pi/p_{cm}^2$ at $T=0$ is also shown.}
\label{fig:sigconst}
\end{figure}

Fig.~\ref{fig:sigconst}
shows 
the cross sections for $\pi\pi$ scattering, 
which are calculated for a 
constant $\left|V_{\rho\pi\pi} \right|=2.4$.
%using the parameter set 
%$m_0=4$~MeV, $\Lambda=665$~MeV, $G_1\Lambda^2=2.5$, and
%$G_2\Lambda^2=1.5$.
%For these parameters 
At $T=0$, we find that the $\rho$-pole  is located
at ${\rm Re}(m_\rho)\simeq 815$~MeV and 
${\rm Im}(m_\rho)\simeq -86$~MeV. This $\rho$-pole corresponds to the
peak of the cross section at $T=0$ in Fig.~\ref{fig:sigconst} with
a width $\Gamma\sim -2{\rm Im}(m_\rho)\sim 170$~MeV. 
As expected, the cross section at $T=0$ touches the unitarity bound.
In comparison with the cross section at $T=0$, 
the peak of the cross section at $T=150$~MeV becomes lower, 
but its position does not shift much.
While the cross sections at $T=0$ and 150~MeV have resonance-like
shapes and can be directly related to the corresponding $\rho$-poles given in 
Fig.~\ref{fig:sigconst}, the relationship between the $\rho$-pole
position and the shape of the cross section at $T=200$~MeV is less clear.
In the latter case, the $\rho$-pole on the 
$\left[1^{(q)},2^{(\pi)} \right]$ sheet is found to be slightly
above the $q\bar q$ 
threshold that is indicated in Fig.~\ref{fig:rhopole}. One may expect in this case,
the shape of the cross section is a combination of the effects
of the $q\bar q$ threshold and the $\rho$-pole we found.
On the other hand, the cross section at $T=220$~MeV displays mainly
the threshold behavior, known in the literature as the Wigner cusp.
In this case the $\rho$-pole shown in Fig.~\ref{fig:rhopole} locates on
the $\left[2^{(q)},1^{(\pi)} \right]$ sheet, and is so far away from
the physical sheet that it cannot make its existence dominant  in the shape
of the cross section.

%In Fig.~\ref{fig:rhot} we show the $\rho$-pole trajectory in the complex
%plane as a function of the temperature for a fixed value of the model
%parameters.
%According to  Fig.~\ref{fig:rhot} 
%for temperatures $T\leq 115$~MeV, the $\rho$-pole position
%depends weakly on
%temperature, while for temperatures $T>115$~MeV, 
%%${\rm Re}(m_\rho)$ becomes smaller, and $\rho$-meson width
%both real and imaginary parts of $m_\rho$ decrease
%as $T$ increases, until the $\rho$-pole
%arrives at the real axis and moves through the cuts there (above the
%$q\bar q$ threshold) into the $\left[2^{(q)},1^{(\pi)} \right]$ sheet that
%is of less physical significance.

\subsection{Cross sections  for energy dependent $\left|V_{\rho\pi\pi}
\right|$}
\label{sect:signonconst}
\begin{figure}
\epsfig{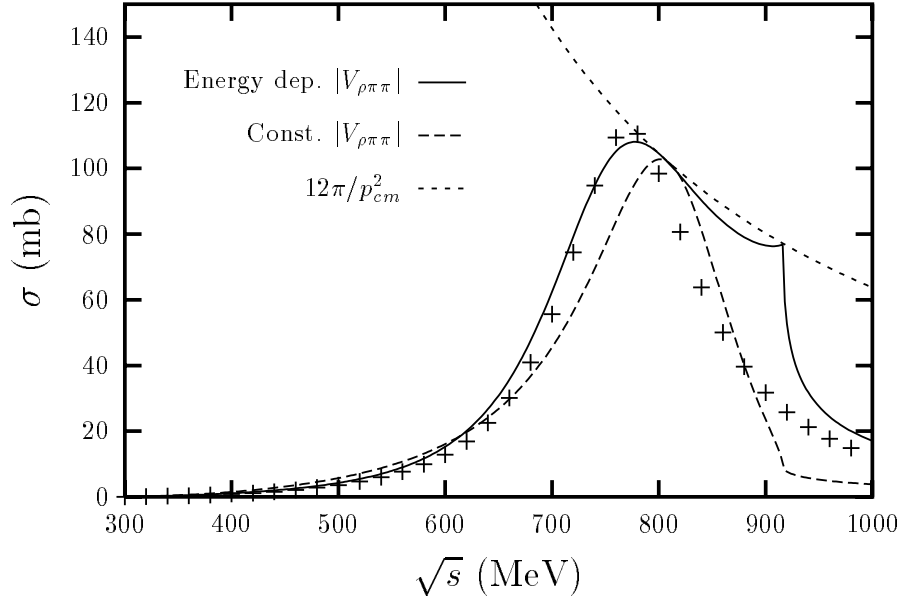}
\bigskip
\caption[]{Energy dependence of the $\pi\pi$ elastic cross section in the 
$\rho$ exchange channel at 
$T=0$ for
two choices of the $\rho\pi\pi$ vertex function (full
energy dependence and approximated by a constant), 
compared to the data (represented by crosses) taken from 
Ref.~\cite{cdFro77}. Also shown is the unitary bound $12\pi/p_{cm}^2$ at 
$T=0$.}
\label{fig:sigfit}
\end{figure}

In our calculations, we have chosen the parameter $G_2$ so that at $T=0$ the
$\rho$-meson appears as a $\pi\pi$ resonance with mass $m_\rho\sim 768$~MeV, 
as one can see from a fit of the cross section $\sigma(\pi^+\pi^-
\rightarrow \pi^+\pi^-)$ given in
Fig.~\ref{fig:sigfit} for an 
energy dependent $\left|V_{\rho\pi\pi} \right|$. 
The data points of $\sigma(\pi^+\pi^-
\rightarrow \pi^+\pi^-)$
translated from the phase shifts of $\pi^+\pi^-$ scattering in the 
vector-isovector channel \cite{cdFro77}, 
instead of the $\pi^+\pi^-$ total
elastic cross section (\eg, given in Ref.~\cite{jpBat70}), 
has been used here in order to 
single out the dominant $\rho$-meson exchange contribution to 
$\sigma(\pi^+\pi^-
\rightarrow \pi^+\pi^-)$ and remove background
contributions  resulting from other channels that are not present in our
calculations.
One sees in Fig.~\ref{fig:sigfit} that 
the cross section 
is properly bounded by the unitarity at $T=0$, and that 
both the position and the shape 
can be reasonably reproduced by the present parameter set.
Fig.~\ref{fig:sigfit} also shows  the  results for 
$\sigma(\pi^+\pi^-
\rightarrow \pi^+\pi^-)$ for a constant 
value of the vertex (
$\left|V_{\rho\pi\pi} \right|=2.4$) and for an energy dependent vertex
$\left|V_{\rho\pi\pi} \right|$. The more realistic choice of the 
energy dependent vertex, with the exception of the area surrounding 
the $q\bar q$ threshold, gives a better 
fit to the experimental cross section, which has been calculated from the
phase shifts $\delta_1^1$ for $\pi\pi$ scattering.
The kink on the right hand side of the curve for energy dependent 
$\left|V_{\rho\pi\pi} \right|$ is due to the opening of the unphysical 
$\rho \rightarrow q\bar q$ channel.
%The good agreement of the two theoretical curves
% justifies the simplification we made 
%towards a constant $\left|V_{\rho\pi\pi} \right|$
%in Sec.~\ref{sect:sigconst}.

\begin{figure}
\epsfig{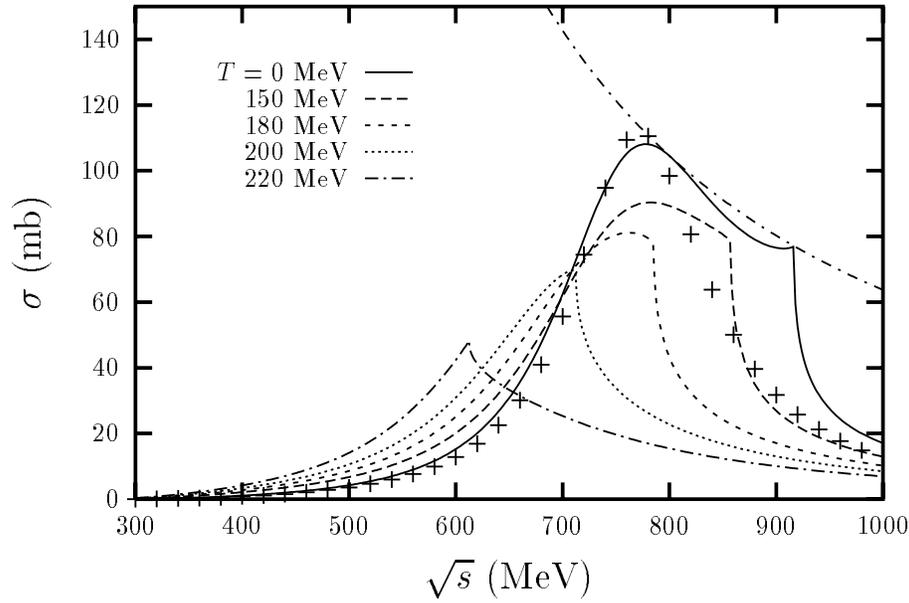}
\bigskip
\caption[]{Elastic $\sigma(\pi^+\pi^-
\rightarrow \pi^+\pi^-)$ at temperatures $T=0$, 150, 180, 200, and 
220~MeV for energy dependent $\left|V_{\rho\pi\pi} \right|$. The
crosses represent the experimental cross section at $T=0$, and the
long dashed line is the unitarity bound $12\pi/p_{cm}^2$ at $T=0$.}
\label{fig:signon}
\end{figure}

Fig.~\ref{fig:signon} illustrates the $\pi^+\pi^-$ elastic cross sections
at different temperatures for energy dependent $\left|V_{\rho\pi\pi} \right|$,
which is entirely determined in Eq.~(\ref{grpp}) by the present model.
These results show analogous features to those for a  constant 
$\left|V_{\rho\pi\pi} \right|$ (Fig.~\ref{fig:sigconst}) 
in that the position of the maximum moves to smaller values of 
$\sqrt{s}$ with increasing temperature. 
One notes that 
the height of the maximum decreases with respect to that shown in 
Fig.~\ref{fig:sigconst} for a constant vertex. This is because
in this case the cross section is always 
dominated by
$q\bar q$ threshold effects, displaying  Wigner cusps. 

%In the temperature region $T\leq 150$~MeV 
%the elastic $\pi\pi$ cross section has a rather weak temperature dependence,
%and for $T\leq 180$~MeV its $T$-dependence reflects
%the $T$-dependence of the (real) $\rho$ (Fig.~\ref{fig:mrhot}).

\section{Results and Conclusions}
\label{sect:concl}
To summarize this work, we have  studied the $\rho$-meson propagator and 
 $\pi\pi$ scattering
via $s$-channel $\rho$-meson exchange in the framework of the extended
NJL-model. In particular, we have argued that 
in addition to the quark loop contribution,
the $\rho$-meson self-energy should include a pion loop contribution
from physical considerations 
which is found to be essential in order to preserve unitarity of the 
$S$-matrix and to
 reproduce the experimental
$\pi\pi$ phase shifts at zero temperature.
In order to maintain current conservation, other diagrams of the same
order in $1/N_c$ expansion must be included. These diagrams reduce to 
the well-known tadpole diagrams of the hadronic model \cite{gal91} in 
the limit in which the quark loops are contracted to a point.

The new part of our work is  a careful study
of the analytic structure of the two-channel
 scattering amplitude.
We have studied the $\rho$-pole 
trajectories in the twofold cut complex energy plane
with respect to the variations of the coupling strength
and temperature.
% and the establishment of 
%clear connections between the $\rho$-pole
%positions and the shapes of the $\pi\pi$ cross section. 
In order to reproduce 
the data of $\pi\pi$
phase shifts at zero temperature, it is required that the unphysical $q\bar q$
threshold should be put above the physical $\rho$ mass, 
and the  $\rho$-meson can be consequently identified as a $\pi\pi$ 
resonance below the $q\bar q$ threshold.
At temperatures $T<215$~MeV,
the $\rho$-meson appears
as a complex pole close to the physical sheet.
Specifically, as the temperature increases, for temperatures $T\leq 130$~MeV
the $\rho$-meson mass increases only slightly and its width is almost
unchanged, while for temperatures $215 \geq T \geq 130$~MeV, both 
the mass and width of the $\rho$-meson decrease. 
For these temperatures
the $\pi\pi$ cross section has a resonance-like shape, and its 
temperature dependence is rather weak and is mainly determined
by the temperature-dependent $\rho$-meson mass and width.
At higher temperatures ($T>215$~MeV), 
however, the poles locate on the unphysical
sheet and $q\bar q$ threshold effects determine the shape of the cross section.
The rather weak dependence of the position and width of the $\rho$-pole
as a function temperature observed in this work
is in agreement with the results of lattice calculations \cite{boy95}.

\acknowledgments
We thank the referee for pointing out the importance of current 
conservation when including the pion contributions to the 
$\rho$-meson polarization functions.
This work has been supported in part by the Deutsche Forschungsgemeinschaft
DFG under the contract number Hu 233/4-4, and by the German Ministry for Education
and Research (BMBF) under contract number 06 HD 742.
\appendix

\section{One-loop polarization functions}
\label{sect:loop}
In the imaginary time formalism for finite temperature, 
one loop polarization functions for different meson modes are given by
\be
-i\Pi^{VV}_{\mu\nu,ab}(q)&=&
\intk 
(-1) {\rm Tr} [ (\gamma_\mu \tau_a) iS(k)(\gamma_\nu
\tau_b) iS(k+q)],
\\
-i\Pi^{PP}_{ab}(q)&=&
\intk 
(-1) {\rm Tr}[(i\gamma_5\tau_a)
iS(k)(i\gamma_5\tau_b)iS(k+q)],
\\
-i\Pi^{PA}_{\mu,ab}(q)&=&
\intk (-1) {\rm Tr} [(i\gamma_5\tau_a)
iS(k)(\gamma_\mu \gamma_5\tau_b) iS(k+q)],
\\
-i\Pi^{AP}_{\mu,ab}(q)&=&
\intk (-1) {\rm Tr} [ (\gamma_\mu \gamma_5\tau_a) iS(k)(i\gamma_5\tau_b)
iS(k+q)],
\\
-i\Pi^{AA}_{\mu\nu,ab}(q)&=&
\intk (-1) {\rm Tr} [ (\gamma_\mu \gamma_5\tau_a) iS(k)(\gamma_\nu
\gamma_5\tau_b) iS(k+q)].
\ee
Here Tr[$\cdots$] is a trace over the color, flavor and spinor space,
$\beta$ is the inverse temperature, 
and the Matsubara sums on $n$ run over the fermionic frequencies
$\omega_n=(2n+1)\pi/\beta$, with $n=0, \pm 1, \pm 2, \cdots$,
which through $k=(i\omega_n,{\bf k})$
enter the quark propagator $S(k)=(\not\! k +m_q)/(k^2-m_q^2)=
[(i\omega_n)\gamma_0-{\bf k\cdot \gamma}+m_q)/[(i\omega_n)^2-E_k^2]$,
with $E_k^2={\bf k}^2+m_q^2$.
After performance of 
the Matsubara summation the limit $\eta \rightarrow 0$ is to be taken, 
and an analytic continuation is to be
applied to 
the boson momentum $q=(i\mu_m,{\bf q})$, $\mu_m=2m\pi/\beta$, 
$m=0, \pm 1, \pm 2, \cdots$, which implies $i\mu_m\rightarrow 
q_0$.
%+i\epsilon$, with $\epsilon\rightarrow 0^+$.
The three-momentum cut-off $\Lambda$ is used to regularize
the divergences of the integrals.

These polarization functions can be decomposed according to their flavor and
spinor structure as,
%\begin{mathletters}
\begin{eqnarray}
&\Pi^{VV}_{\mu\nu,ab}(q^2)&=
\left[\Pi^{VV}_T(q)T_{\mu\nu}+\Pi^{VV}_L(q)L_{\mu\nu}
\right]\delta_{ab},
\label{e:appVV}
\\          
&\Pi^{AA}_{\mu\nu,ab}(q^2) &=[\Pi^{AA}_T(q^2) 
T_{\mu\nu}+\Pi^{AA}_L(q^2) L_{\mu\nu}]\delta_{ab},
\label{e:appAA}
\\
&\Pi^{AP}_{\mu,ab}(q^2) &=\Pi^{AP}(q^2) \hat{q}_\mu \delta_{ab},
\qquad \hat{q}_\mu=\frac{q_\mu}{\sqrt{q^2}},
\label{e:appAP}
\\
&\Pi^{PP}_{ab}(q^2)&=\Pi^{PP}(q^2)\delta_{ab},
\label{e:appPP}
\end{eqnarray}
%\end{mathletters}
where the standard longitudinal and transverse tensors are
$L_{\mu\nu}=\hat{q}_\mu\hat{q}_\nu$ and $T_{\mu\nu}=g_{\mu\nu}-
\hat{q}_\mu\hat{q}_\nu$.

In terms of  the following elementary integrals
\bem
\be
I_1 &=&
\intk {1 \over k^2-m_q^2},
\label{e:I1}
\\
I_2(p) &=&
\intk {1 \over [k^2-m_q^2]\;[(k+p)^2-m_q^2]},
\label{e:I2}
\\
I_3(p^\prime,p) &=&
\intk {1 \over (k^2-m_q^2)\;[(k+p^\prime)^2-m_q^2]\;[(k+p)^2-m_q^2]},
\label{e:I3}
\ee
\eem
one-loop polarization functions can be evaluated as
\be
\Pi^{VV}_T(q^2)&=&
{4\over 3}iN_fN_c[(q^2+2m_q^2)I_2(q)-2I_1],
\label{e:a7}
\\
\Pi^{VV}_L(q^2)&=& 0,
\label{e:a8}
\\
\Pi^{PP}(q^2) &=&
4iN_fN_c I_1 -2iq^2N_fN_c I_2(q),
\\
\Pi^{PA}(q^2)&=&
4m_qN_fN_c I_2(q)/\sqrt{q^2},
\\
\Pi^{AP}(q^2)&=&
-4m_qN_fN_c I_2(q)/\sqrt{q^2},
\\
\Pi^{AA}_L(q^2)&=&
-8im_q^2N_fN_cI_2(q),
\\
\Pi^{AA}_T(q^2)&=&
\Pi^{VV}_T(q^2) +\Pi^{AA}_L(q^2).
\ee

%%%%%%%%%%%%%%%%%%%%%%%%%%%%%%%%%%%%%%%%%%%%%%%%%%%%%%%%%%%%%%%%%%%

%%%%%%%%%%%%%%%%%%%%%%%%%%%%%%%%%%%%%%%%%%%%
\newpage
%
% figure 1
%

%
% figure 2
%

%
% figure 3
%
\newpage

%
% figure 4
%

%
% figure 5
%

%
% figure 6
%

%
% figure 7
%

%
% figure 8
%

%
% figure 9
%

%
%figure 10
%

%
% figure 11
%

%
% figure 12
%

\end{document}